\documentclass{amsart}
\usepackage{amsaddr}

\usepackage{amsthm}
\usepackage{amsmath}
\usepackage{natbib}
\usepackage[colorlinks,citecolor=blue,urlcolor=blue,filecolor=blue,backref=page]{hyperref}
\usepackage{graphicx}

\usepackage{amsfonts}
\usepackage{mathtools}
\usepackage{dsfont}
\usepackage[algo2e]{algorithm2e}

\numberwithin{equation}{section}
\theoremstyle{plain}
\newtheorem{theorem}{Theorem}[section]
\newtheorem{lemma}[theorem]{Lemma} 
\theoremstyle{remark}
\newtheorem*{remark}{Remark}

\DeclareMathOperator{\diag}{dg}
\DeclareMathOperator{\rank}{rank}
\graphicspath{{figures/}}
\DeclareGraphicsExtensions{.pdf, .png}
\newcommand{\reffig}[1]{{Figure~\ref{#1}}}

\newcommand{\refsec}[1]{{Section~\ref{#1}}}
\newcommand{\reftab}[1]{{Table~\ref{#1}}}

\newcommand{\refalg}[1]{{Algorithm~\ref{#1}}}

\newcommand{\pr}{\mathbb P}

\newcommand{\bb}{\mathbb}

\newcommand{\ind}{\operatorname{\mathds{1}}}

\DeclareMathOperator*{\tr}{Tr}

\DeclareMathOperator*{\argmin}{arg\,min\;}

\title{Shuffled linear regression through graduated convex relaxation}

\author{Efe Onaran}
\address[EO]{Viterbi Faculty of Electrical and Computer Engineering, Technion - Israel Institute of Technology}
\email{efeonaran@campus.technion.ac.il}

\author{Soledad Villar}%
\address[SV]{Department of Applied Mathematics \& Statistics, and Mathematical Institute for Data Science, Johns Hopkins University }
\email{soledad.villar@jhu.edu}

\begin{document}
\maketitle
\begin{abstract}
The shuffled linear regression problem aims to recover linear relationships in datasets where the correspondence between input and output is unknown. This problem arises in a wide range of applications including survey data, in which one needs to decide whether the anonymity of the responses can be preserved while uncovering significant statistical connections. In this work, we propose a novel optimization algorithm for shuffled linear regression based on a posterior-maximizing objective function assuming Gaussian noise prior. We compare and contrast our approach with existing methods on synthetic and real data. We show that our approach performs competitively while achieving empirical running-time improvements. Furthermore, we demonstrate that our algorithm is able to utilize the side information in the form of \emph{seeds}, which recently came to prominence in related problems.
\end{abstract}


\section{Introduction}\label{intsec}

Shuffled linear regression is a weakly-supervised variant of the commonly studied linear regression problem. In classical linear regression, the focus is 
on uncovering significant relationships between two sample sequences $X = \{ x_1, x_2, \ldots x_n\}$ and $Y = \{ y_1, y_2, \ldots y_n\}$, $x_i \in \mathbb{R}^{d_x}$ and $y_i \in \mathbb{R}^{d_y}$, which contain two different kinds of information about a common set of observational entities. In contrast, shuffled linear regression sets to recover the same information when there is no correspondence between the entities representing samples in $X$ and $Y$. In other words, the sample sequences, $X$ and $Y$, are assumed to have lost the ordering in the data acquisition process. This setting arises naturally in a wide range of applications in the social sciences \cite{hejblum2019probabilistic,steorts2016bayesian}, computer vision \cite{collier2016minimax,lowe2004distinctive,crandall2006weakly}, computational biology \cite{baron2019cell} and healthcare \cite{li2004protection}. 

Broadly, the task aims to aggregate and relate data from multiple sources when the identification of records is either unavailable, missing or undesirable. For example, in computational biology, one might be interested in relating the presence of specific surface markers on individual cells with their respective intra-cellular molecular information given by their gene expression. In applications such as  fluorescence-activated cell sorting \cite{herzenberg2006interpreting,baron2019cell} such information is not available due to experimental limitations. Light scattering experiments can help characterize a small number of markers that the cells can exhibit on their surface. RNA-sequencing can then characterize the genomic readout or gene expression of cells. However, the light scattering and the RNA-sequencing experiments can not be performed concurrently, which causes the loss of correspondence between two sets of measurements.

Another relevant use-case of shuffled learning is in nano-scale communications. Due to very high coupling (interference among transmitters that aim to communicate with a common receiver) in the molecule level \cite{rose2014signaling}, the reliable communication heavily depends on the identifying signals, called pilot signals, each transmitter has to add to the beginning of its message. With increasing size of the network, the overhead due to the pilot signals gets unmanageably large. One idea to overcome this ever increasing overhead is to omit the pilot signals completely which leads to the unknown shuffling inherent in the nature of any learning task in this type of communication network.  

Finally, in some scenarios, the correspondence between labels and features is intentionally hidden. When dealing with sensitive data, such as in healthcare, it is a standard practice to shuffle the data before its publication to keep the anonymity of the patients \cite{li2004protection,li2015ensuring}. There is a growing body of work under the umbrella term \emph{de-anonymization} originating from the observation that shuffling data is not enough to maintain statistical confidentiality in all cases \cite{narayanan2008robust}. How to make sure that the anonymity of users is protected while preserving the statistical information contained in the data is an important question that has recently attracted significant research efforts \cite{bonawitz2017practical,cai2021cost}.

\subsection{Contribution}
We propose an algorithm for shuffled regression by leveraging connections with an optimization problem over the set of permutations called \emph{seriation}, which we detail in \refsec{seriation_sec}. We perform numerical experiments which explore the efficiency-accuracy trade-off scale of our algorithm and other selected benchmark algorithms. In particular, we present experimental results on both synthetic shuffled data generated through a linear regression model with additive noise, as well as real datasets, that serve as empirical evidence to the following main observations:

\begin{itemize}
    
     \item Our algorithm provides a fast output without significant sacrifice in accuracy in the generative model. In real datasets, our algorithm performed both faster and more reliable linear regression than benchmark methods. 
     

      \item Our algorithm seamlessly allows the inclusion of side information in the form of partial correspondence, thus improving performance and mitigating overfitting in real datasets, when available. This is a valuable asset to our algorithm especially considering overfitting has been observed to be hopelessly unavoidable in most real datasets.

    
\end{itemize}

\subsection{Notation}
Throughout this manuscript, the entry at the $i$'th row and $j$'th column of a matrix $A$ is referred to as $A_{ij}$. $\|A\|_F$  or $\|A\|$ denotes the Frobenius norm, namely the 2-norm of the vectorized version of the matrix: $\|A\|_F= \|A\|= \sqrt{\sum_{i,j} A_{ij}^2}$. $\langle A,B\rangle$ stands for the matrix inner product of $A$ and $B$ that is $\langle A,B\rangle=\tr(A^\top B).$ The $i$'th smallest eigenvalue of the matrix $A$ is denoted as $\lambda_i(A)$. The operator $\diag(\cdot)$ returns the input matrix with all its non-diagonal entries masked (replaced with zeros).

$\mathcal{P}^n$ denotes the set of $n\times n$ permutation matrices and we sometimes omit the size in the superscript when it is clear from the context. $\mathbf{1}$ is the all-one column vector of length clarified by the context it is used. $\mathbf{I}$ is the identity matrix again of appropriate size, occasionally specified as a subscript to help the reader. To ease the notation throughout this manuscript, we think of permutation matrices \emph{modulo transpose} meaning we use $\Pi$ for both the permutation matrix $\Pi$ and its transpose, when it is clear which one we refer to from the context. Lastly, $A\sim\mathcal N(\mu, \Sigma)$ means the entries of the matrix $A$ are multivariate Gaussian distributed with mean $\mu$ and covariance matrix $\Sigma$.

\subsection{Paper organization}
In \refsec{sec:setting}, we formally introduce the problem setting and provide an overview of existing work. In \refsec{sec:optimization}, we introduce our approach and detail it as an optimization problem. In \refsec{sec:experiments}, we provide extensive numerical experiments and compare our algorithm to existing approaches across a variety of metrics. Our concluding remarks are in \refsec{sec:discussion}.

\section{Problem Setting}
\label{sec:setting}

Our framework of analysis of the shuffled linear regression is based on a Bayesian generative model for unpaired data $(X,Y)$, where the correspondence between the independent variables $X$ and the linearly dependent observations $Y$ is unknown. The model can be more concretely written as 
\begin{equation}
    Y= \Pi^{true} X \beta +\epsilon, \label{gaus}
\end{equation}
where $X$ is a real-valued matrix of $n$ samples of dimension $d_x$ (i.e. $X \in \mathbb{R}^{n \times d_x}$) and $Y$ is a matrix of corresponding outputs ($Y \in \mathbb{R}^{n \times d_y}$) that has been shuffled by an unknown permutation matrix $\Pi^{true}$, and obfuscated by an additive noise vector $\epsilon \in \mathbb{R}^{n \times d_y}$. In particular, $\Pi^{true}$ is an $n\times n$ binary matrix with a unique non-zero entry in each row and column, and is assumed to hold uniform a priori distribution among the set of such permutation matrices. We also assume the entries of $\epsilon$ are i.i.d. with zero mean, therefore we do not statistically distinguish between $\epsilon$ and any shuffled version of it.

Given the features $X$ and the shuffled labels $Y$, three parallel problems can be stated. First, seeking the recovery of the regression coefficients $\beta \in \mathbb{R}^{d_x \times d_y}$, which would correspond to \emph{shuffled learning} problem that we have described previously. Second, recovering the planted permutation $\Pi^{true}$. Recovering $\Pi^{true}$ is important for applications that aim for identification of the samples, while making $\Pi^{true}$ irrecoverable is the goal for applications where the identification of the sources is intentionally hidden. Third, reducing the error between the proposed model, composed of estimates of both $\beta$ and $\Pi^{true}$, and the observed labels or \emph{training error}, which we call \emph{denoising}.

In this paper, we consider and present results regarding all three forms of recovery (or learning) tasks mentioned above. We further evaluate the impact prior knowledge has on this inference task by considering a \emph{seeded} version of the problem, where the correspondence between some pairs of rows of $X$ and $Y$ is known. In other words, the seeded scenario corresponds to the case where some non-zero entry locations of the permutation matrix $\Pi^{true}$ are given as partial information.   

\subsection{Related Work}

Historically related to the \emph{broken samples problem} \cite{goel1975re}, shuffled linear regression has received considerable revived interest in the last five years from both theoretical and practical viewpoints, mainly motivated by the applications we pointed out in \refsec{intsec}.
  In \cite{unnikrishnan2018unlabeled}, the authors considered the noise-free and one dimensional output ($d_y=1$) version of the problem motivated by the robotics task of  simultaneous localization and
  mapping, where a robot is sensing an unknown environment, without the knowledge of its own spatial location \cite{stachniss2016simultaneous}. In this noiseless setting, where the entries of the feature matrix $X$ are i.i.d and continuous, the authors of \cite{unnikrishnan2018unlabeled} proved that $\Pi^{true}$ can be recovered   
  with probability $1$ if and only if $n\geq 2d_x$. However, no efficient algorithm for 
  recovering either $\beta$ or $\Pi^{true}$ was provided. In fact, finding the \emph{maximum likelihood} estimate of $\Pi^{true}$, denoted $\Pi^{ML}$, given $X$ and $Y$ under the assumption of i.i.d. standard Gaussian entries for $X$ and i.i.d. Gaussian noise entries, is NP-hard for any multi-dimensional input case ($d_x \geq 2 $)  as proven in \cite{pananjady2017linear}, where the authors also established a statistical bound for the perfect recovery of the planted permutation ($\Pi^{true}$) in the noisy and  one-dimensional output ($d_y=1$) case. A multi-dimensional output extension of this result in \cite{zhang2019permutation} established that 
\begin{align}
\pr[\Pi^{ML} = \Pi^{true}] \to 1 \quad &\text{if and only if}    \quad  \frac{\log(1+ \mathrm{snr})}{\log(n) }  > C_1 \label{mltrue}\\
&\text{where}\quad 
  \mathrm{snr}  = \frac{\|\beta \|_F^2}{\sigma^2} \nonumber.
\end{align}
  where $C_1$ is a constant empirically estimated to be between 3 and 4. In \eqref{mltrue} the standard deviation of i.i.d. Gaussian entries of the noise vector $\epsilon$ is denoted by $\sigma$. Furthermore, when $d_y=1$ any algorithm needs $\mathrm{snr}>C_2\cdot d_x/{\log \log n}$ for the approximate recovery (in $\ell_2$ sense) of $\beta$ \cite{hsu2017linear}. 

Complementing the information theoretic treatment of the problem, empirical approaches to shuffled regression exist, very few with theoretical guarantees. Alternating optimization techniques were proposed in \cite{haghighatshoar2017signal} but observed to be heavily dependent on a good choice of initialization. In \cite{abid2017linear}, the authors propose an algorithm based on the permutation invariance of the self-moments of features and labels. The resulting algorithm minimizes the a posteriori difference between the self-moments of feature columns and those of the labels up to moments of degree $d_x$. In this process, the regression coefficients, but not the underlying permutation, are recovered. 
However, this algorithm becomes computationally infeasible for a large number of features, $d_x$, due to its expensive computation of moments. Another caveat is that the algorithm is not easily generalizable to the case $d_y>1$, requiring additional heuristics. Recently, some theoretical guarantees for a generalized version of the self-moment idea, \emph{symmetric (permutation-invariant) polynomials}, was obtained in \cite{tsakiris2020algebraic} for the noise-less case. Branch-and-bound methods for the concave formulation \cite{peng2020linear} of the problem also seem to suffer from infeasible computational complexity for dimensions, $d_x$, as small as 9.
In addition to these, a spectral algorithm based on the \emph{leverage scores} of $Y$ and $X$ was suggested in \cite{pananjady2017denoising} for the noiseless case and in the restrictive setting of $\rank(X)\leq \rank(\beta)$. The leverage score is defined as $l(X) \coloneqq \diag(U_X U_X^\top)$ where 
\begin{align}
    X = U_X\Sigma_X V_X^\top \nonumber
\end{align}
is the reduced singular value decomposition of $X$.  

We also note that shuffled regression is a particular case of uncoupled isotonic regression, where the goal is to learn an underlying monotonic relationship  between the input and output samples  in the form of $y_i = f(x_i) + \epsilon_i$,
where the coupling between the pairs $(x_i,y_i)$ is lost \cite{rigollet2019uncoupled,carpentier2016learning}. However, the optimal transport tools developed for uncoupled isotonic regression are unfortunately restricted to  one dimensional inputs and outputs. The monotonicity constraint is not amenable to high-dimensional data, thus the framework is not immediately applicable 
to the general setting with arbitrary input and output dimensions considered in this paper.

Finally, our work is strongly motivated by the recent line of research efforts that involve optimization over permutation matrices which arise in several applications distinct from shuffled regression. Specifically, when a problem  poses a loss function which is quadratic in a permutation matrix, the problem is called \emph{Quadratic Assignment Problem (QAP)}. The NP-hardness of QAP was proven earlier \cite{sahni1976p}. However, as with other applications in data science, the average case complexity of QAP has recently attracted interest.  \emph{Graph matching}, the problem of permuting the node sets of two graphs such that edge alignment is maximized,  is a special case of QAP and has been actively explored through the lens of both statistical limits \cite{onaran2016optimal,cullina2016improved}, and that of existence of efficient algorithms \cite{lyzinski2016graph},  \cite{onaran2017projected}. 


Remarkably, the use of seeds (few pairs of given matched nodes corresponding to ground truth) in Bayesian models for graph matching has proven crucial. Several practical algorithms have been proposed that successfully leverage seed information \cite{lyzinski2014seeded,shirani2017seeded,kazemi2015growing,yartseva2013performance}. This naturally rises the question regarding the possible use of seeds in other special cases of QAP such as in shuffled regression.
While the use of partial side information was studied in shuffled regression, \cite{slawski2019linear, slawski2020sparse, slawski2020two} where the planted permutation was given to be in the close \emph{Hamming} vicinity of a given permutation, or the case where a block diagonal structure was assumed on $\Pi^{true}$ \cite{shi2021spherical}; to the best of our knowledge, there is no work that studies the effect of seeds or algorithm that exploits the seed information in the shuffled regression context.

\section{Optimization approaches to shuffled linear regression}
\label{sec:optimization}
\subsection{Shuffled linear regression as optimization over permutations}

Given our model \eqref{gaus}, assuming Gaussian i.i.d. noise with zero mean, and uniformly distributed shuffling matrix over the set of permutation matrices of size $n\times n$, it is straightforward to observe that recovering the maximum-likelihood solutions for the planted linear coefficients and the hidden permutation is equivalent to solving the following optimization problem:
\begin{equation}\label{opti}
     \argmin_{\beta, \Pi} \| \Pi Y - X \beta \|_F^2. 
\end{equation}

In order to mitigate possible ill-conditioning, we consider the ridge regression version of this optimization problem with Tikhonov regularization,
\begin{equation}
    (\beta^*, \Pi^*)_\lambda = \argmin_{\beta, \Pi} \|\Pi Y - X \beta \|_F^2 + \lambda\|\beta\|_F^2. \label{eq.problem}
\end{equation}
The optimal choice of $\lambda$, which we do not delve into in this paper, depends on the variance (assumed unknown) of each entry of the noise vector $\epsilon$. Note that given $\Pi^*$, the problem amounts to solving an ordinary linear regression,
\begin{equation}
    \beta_{\Pi^*,\lambda} = \left( X^\top X + \lambda \mathbf{I}_{d_x} \right)^{-1}X^\top \Pi^* Y. \label{1241}
\end{equation}
Hence, we can reduce \eqref{eq.problem} to an optimization over the set of permutation matrices,
\begin{alignat}{2}
   \Pi^* &= \argmin_{\Pi} && \left( \Pi Y - X \beta_{\Pi,\lambda}  \right)^\top \left( \Pi Y - X \beta_{\Pi,\lambda}   \right) + \lambda\|\beta_{\Pi,\lambda}\|_F^2 \nonumber\\
    &= \argmin_{\Pi} && Y^\top \Pi^\top \Pi Y - Y^\top \Pi ^\top X \beta_{\Pi,\lambda} - \beta_{\Pi,\lambda}^\top X^\top \Pi Y \nonumber\\
    & &&+\beta_{\Pi,\lambda}^\top X^\top X \beta_{\Pi,\lambda} +\lambda\beta_{\Pi,\lambda}^\top \beta_{\Pi,\lambda}. \nonumber
\end{alignat}

To simplify the notation, we denote $M= \left( X^\top X + \lambda \mathbf{I}_{d_x} \right)^{-1}$ and $S = X M X^{\top}$, and we note that both are symmetric matrices that can be explicitly computed from the data. Then replacing the closed form optimal $\beta$ in \eqref{1241} yields to
\begin{alignat}{2}
    {\Pi}^* &= \argmin_{\Pi} &&
     Y^\top \Pi^\top \Pi Y - Y^\top \Pi^\top S \Pi Y - Y^\top \Pi^\top S^\top \Pi Y \nonumber\\
     & && +Y^\top \Pi^\top SS^\top \Pi Y  + \lambda Y^\top \Pi^\top XM^\top M X^\top \Pi Y \nonumber \\
     &=\argmin_{\Pi} && Y^\top \Pi^\top \left[\left( S- \mathbf{I}_n\right)^2 + \lambda X M^\top M X^\top \right] \Pi Y. \label{eq:opt}
\end{alignat}
which implies that finding the maximum likelihood solution to the shuffled linear regression with Gaussian prior on the coefficients and uniform prior on the shuffling is essentially equivalent to minimizing a quadratic function \eqref{eq:opt} of permutation matrices. Given $\Pi^*$, estimate for $\beta$ can be easily obtained through the closed form expression in \eqref{1241}.

\subsection{Shuffled linear regression as generalized seriation}
\label{seriation_sec}
Here, we derive an algorithm for shuffled regression inspired from a successful computational approach towards a related problem called \emph{seriation}. 

The seriation problem seeks to reconstruct a linear order between variables using unsorted, pairwise similarity information. The problem originates from archaeology, as a method to analyze artifacts from different archaeological sites to order the sites chronologically \cite{robinson1951method}. Seriation is also an important problem in genomics \cite{recanati2018robust} where the goal is to reconstruct a DNA strand from random overlapping sub-samples using pairwise similarity scores (degrees of overlap).

The most general formulation of the seriation problem takes an $n\times n$ similarity matrix $A$ as input and searches for a re-ordering of rows and columns such that $A_{ik}\leq \min(A_{ij}, A_{jk})$ for all $i,j,k$ with $1\leq i\leq j \leq k \leq n$.

\cite{fogel2013convex} shows that the seriation problem over a certain class of similarity matrices is equivalent to an optimization problem over permutation matrices called the 2-SUM problem, that takes the following form:
\begin{align}
    \min_{\pi \in \mathcal{P}\mathcal{H}^n} \sum_{i=1}^n \sum_{j=1}^n A_{ij}(\pi(i)- \pi(j))^2,\label{240}
\end{align}
where $\mathcal{P}\mathcal{H}^n$ denotes the set of permutahedrons, or permutation vectors, which is the set,
\[\mathcal{P}\mathcal{H}^n = \{\Pi \mathbf e\; : \Pi \in \mathcal{P}^n\}   \]
\noindent where $\mathbf e = (1, 2,\ldots,n)^\top$. Problem~\eqref{240} is equivalent to the formulation,
\begin{align}
    \min_{\pi \in \mathcal{PH}^n} \pi^\top L_A \pi, 
    \label{seriation}
\end{align}

\noindent where $L_A$ is the Laplacian of $A$, namely $L_A = \diag(A \mathbf{1}) - A$. 

The optimization problem \eqref{seriation} is known to be NP-hard \cite{george1997analysis}. Therefore a series of practical relaxations have been proposed. For instance, a spectral approach using the Fiedler vector (eigenvector corresponding to the second smallest eigenvalue) \cite{atkins1998spectral}; and a convex relaxation to the space of Birkhoff Polytope \cite{fogel2013convex},
 that is the set of double stochastic matrices, which we will denote as~$\mathcal{D}$,
\[\mathcal{D}^n \coloneqq \{D\in\mathbb{R}^{n\times n}\;:\; \mathbf{1}^\top D = \mathbf{1}^\top,\; D\mathbf{1} = \mathbf{1},\; D_{ij}\geq 0\}.\]

Although the relaxation to Birkhoff Polytope obviates the combinatorial nature of the problem by convexifying the search set, it introduces more variables to keep track of, namely the $n^2$ entries of the double stochastic matrix. A convex relaxation via a more concise representation of permutations called \emph{sorting networks}, which reduces the number of tracked variables and helps cutting the computational load, was considered in \cite{goemans2015smallest},  \cite{lim2014beyond}. Finally, the algorithm proposed in \cite{evangelopoulos2018continuation} brings the best of both worlds together by relaxing the search set to the convex polytope of the permutahedron denoted $\mathcal{DH}^n$, that is
\[\mathcal{DH}^n=\{ D\mathbf{e}\; : \; D\in\mathcal{D}^n \}.\]

Essentially, the relaxation considered in \cite{evangelopoulos2018continuation} can be concisely expressed as follows
\begin{align}
    \min_{\pi \in \mathcal{DH}^n} \pi^\top L_A \pi.\label{rel}  
\end{align}
As it is though, the relaxation in \eqref{rel} suffers from the fact that the center of the polytope, that is 
\[\frac{1}{n}\mathbf{1}\mathbf{1}^\top,\]
results in an objective that scales as $O(n^{-1})$, therefore  for large $n$ dominates meaningful solutions. 

 To overcome this behavior and produce a non-trivial solution, a regularized alternative is used, which is quite common in convex relaxations of problems with sparse search sets,
\begin{align}\label{239}
    \min_{\pi \in \mathcal{DH}^n} f_{\mu}(\pi) \coloneqq \pi^\top L_A \pi - \mu \|H\pi \|_F^2,
    \end{align}
where $H= \mathbf{I}_n - \frac{1}{n}\mathbf{1}\mathbf{1}^\top$ is the centering matrix. Note that the last term in \eqref{239} is maximized when $\pi\in \mathcal{PH}^n$. Although its search set is convex, this objective is generally nonconvex, unless the values of $\mu$ considered are such that $L_A - \mu H$ is positive semidefinite. When $\mu \leq \lambda_2(L_A)$ the problem is convex, whereas if $\mu \geq \lambda_n(L_A)$ the problem is concave. Increasing $\mu$ leads to gradual changes in the objective from a convex to an indefinite, and finally to a concave problem, hence the name GnCR (Graduated non-Convexity Relaxation) in \cite{evangelopoulos2018continuation}. The concavity ensures the end result is a vertex of the search set, $\mathcal{DH}^n$, that is a permutation vector. At each iteration, the GnCR algorithm, illustrated in entirety in \refalg{Seriation_Algorithm}, takes the gradient descent step with respect to the objective function $f_{\mu}(\pi)$ followed by a \emph{Frank-Wolfe} search over the line that connects the current solution to the gradient-descent step.

\begin{algorithm2e}[ht]
\caption{Seriation Algorithm (GnCR) of \cite{evangelopoulos2018continuation}}
   \label{Seriation_Algorithm}
\SetAlgoLined
    \KwIn{Laplacian matrix $L_A$, continuation parameter $\gamma > 1$}
     
\nl    $\mu \leftarrow \mu_0; \; \pi \leftarrow \pi_0$

    \While{$\mu \leq \lambda_n(L_A)$}{
    \While{not converged}{
    \nl    $ \pi^* \leftarrow \arg \min_{\pi^{next} \in \mathcal{DH}^n } \langle \nabla f_{\mu} (\pi), \pi^{next} \rangle\;$  \tcp{ \textcolor{magenta}{ Gradient-Descent step} }
     \nl    $\alpha \leftarrow \arg \min_{\alpha \in [0,1]} f_{\mu} (\alpha \pi^* + (1-\alpha) \pi)\;$ \tcp{ \textcolor{magenta}{Frank-Wolfe search} }
    \nl     $\pi \leftarrow \alpha \pi^* + (1-\alpha) \pi$}
     \nl $\mu \leftarrow \gamma \mu$}
     \KwOut{Order of the final solution $\pi$}
\end{algorithm2e}

We emphasize that the solution of Step-2 in \refalg{Seriation_Algorithm}, $\pi^*$, is a permutation. This is due to the fact that a bounded linear program is optimized at a vertex of the constraint set. Thus, $\pi^*$ can be easily computed using the Hardy - Littlewood - P\'olya \emph{rearrangement inequality} \cite{hardy1952inequalities}, which states that two vectors $a$ and $b$ assume the minimum shuffled inner product when sorted in opposite orders.

\subsection{GnCR for Shuffled Regression with Seeds}
\label{gncrsec}
In this section, we employ the GnCR idea described previously for the seriation problem, to the shuffled regression setting. We further generalize the algorithm to take advantage of the given partial \emph{side information}, where \emph{seeded} correspondences (locations of some non-zero entries of $\Pi^{true}$) are available. We first describe the seeded shuffled regression rigorously and set the notation below. 

\emph{Seeded Shuffled Regression.} 
Let $C\subset [n]$ be a set of indices. The side information associated with a set $C$ is represented by the restriction of the true permutation matrix  $\Pi^{true}$ to $C$. Namely, let $\pi^{seeds}: C\to [n]$ be an injective function with
$\pi^{seeds}(i) = j $ if and only if $ \Pi^{true}_{ij} = 1.$ 
The seeded information can be expressed as pairs $(i,j)$ where $i\in C, j \in C^*$, and  $C^*=\{\pi^{seeds} (i) : i\in C\}\subset[n]$.

We tackle the optimization problem put forth in \eqref{eq:opt} constrained to seed information. The objective function in this case can be written as  
\begin{equation}\label{103}
    \argmin_{\Pi\in C}\;\tr Y^\top \Pi^\top L \Pi Y,
\end{equation}
where, with some abuse of notation, we denote $C$ as the set of permutation matrices that is consistent with $\pi^{seeds}$ and 
\begin{equation*}
    L = \left( S- \mathbf{I}_n\right)^2 + \lambda X M^\top M X^\top .
\end{equation*}

The search variable can be decomposed as  $\Pi = \Pi_C + \Pi_{\overline{C}}$,
where $\overline{C}=[n]\setminus C$ and $\Pi_C$ is an $n$-by-$n$ \emph{partial} permutation matrix that maps $C\to C^*$  and has zero in all other entries. $\Pi_{\overline{C}}: \overline{C}\to [n]\setminus{C^*}$ is defined similarly.  Note that seeds fix $\Pi_C$, therefore the objective function can be rewritten as
\begin{equation}\label{147}
    \argmin_{\Pi_{\overline{C}}}\;2\tr Y^\top  \Pi_C^T L \Pi_{\overline{C} } Y  + \tr Y^\top \Pi_{\overline{C}}^\top  L \Pi_{\overline{C}} Y.
\end{equation}

\begin{remark}
\eqref{147} can be rewritten as
\begin{equation}\label{224}
\argmin_{\Pi_{\overline{C}}}\; g(\Pi_{\overline{C}}) 
\end{equation}
with
\begin{align*}
    g(\Pi_{\overline{C}}) \coloneqq 2\tr  \widetilde{Y}^\top \Pi_C^T \widetilde{L} \Pi_{\overline{C}} \widehat{Y}  + \tr \widehat{Y}^\top \Pi_{\overline{C}}^\top  \widehat{L} \Pi_{\overline{C}} \widehat Y,
\end{align*}
where $\widehat Y$, $\widetilde Y$, $\widehat L$ and $\widetilde L$ are row-sampled versions of $Y$ and $L$, that is
\begin{align*}
\widehat Y &= Y[\overline C^*, :],\\
\widetilde Y &= Y[C^*,:],\\
\widehat L  &= L[\overline C, \overline C], \quad\textrm{and} \\
\widetilde L &= L[C, \overline C], \quad\textrm{in Python notation.}
\end{align*}
Here, $\Pi_C$ and $\Pi_{\overline C}$ are $|C|\times |C|$ and $|\overline C|\times |\overline C|$ full permutation matrices unlike in \eqref{224}. Therefore we reduce the search set to permutation matrices of size $|\overline C|$.  
\end{remark}

Note that the shuffled regression problem with seeds $C$ has been reduced to an optimization problem over the set of permutations of size $|\overline C|=n-|C|$, however the objective \eqref{224} is still implicitly using the side information ($\Pi_C$) to estimate a consistent $\Pi_{\overline C}$, via the first term in \eqref{224}. 

The objective function \eqref{224} is a mix of linear and quadratic terms. In order to (approximately) solve this optimization problem we follow Frank-Wolfe (conditional gradient) steps on the Birkhoff polytope with graduated non-convexity as in \cite{evangelopoulos2018continuation} with the difference that the objective function there was purely quadratic which is the unseeded special case of our more general scenario.   
 
 In order to enforce the permutation constraint on the Birkhoff polytope we add a third term to the objective function in \eqref{224},
 \begin{align}\label{231}
 \argmin_{D \in \mathcal{D}^{|\overline C|}}\; g_\mu(D) &\coloneqq 2\tr  \widetilde{Y}^\top \Pi_C^T \widetilde{L} D \widehat{Y}  + \tr \widehat{Y}^\top D^\top  \widehat{L} D \widehat Y - \mu \| H D \widehat{Y}\|_F^2\\
 &=2\tr  \widetilde{Y}^\top \Pi_C^T \widetilde{L} D \widehat{Y}  + \tr \widehat{Y}^\top D^\top  \widehat{L}_\mu D \widehat Y \nonumber
 \end{align}
where
 \begin{align}
     \widehat{L}_\mu = \widehat{L} - \mu H.
 \end{align}
 
Note that the gradient of the objective function in \eqref{231} satisfies 
\begin{equation*}
\frac{1}{2}\nabla g_\mu(D)^\top = \widehat{Y} \widehat{Y}^\top  D^\top \widehat{L}_{\mu} + \widehat{Y} \widetilde{Y}^\top \Pi_C^T \widetilde{L},
\end{equation*}
 and the gradient descent step, 
\begin{equation*}
\argmin_{D^{next}\in \mathcal D} \tr \nabla g_\mu(D)^\top D^{next},
\end{equation*}
always results in a permutation matrix, which is a vertex of the Birkhoff polytope, as it is an instance of linear programming. The Frank-Wolfe steps for shuffled regression altogether look like
\begin{align}
    \Pi^* &\leftarrow \argmin_{\Pi\in \mathcal P} \tr \left[(\widehat{Y} \widehat{Y}^\top  D^\top \widehat{L}_{\mu} + \widehat{Y} \widetilde{Y}^\top \Pi_C^T \widetilde{L} ) \Pi\right] \nonumber\\
    \alpha &\leftarrow \argmin_{0\leq \alpha\leq 1} g_\mu (\alpha\Pi^* + (1-\alpha) D) \label{406}\\
    D &\leftarrow \alpha \Pi^* + (1-\alpha) D  \nonumber
\end{align}
Note \eqref{406}, where the optimal step is calculated, is the optimization of a quadratic expression and can be written as follows,
\begin{align}
    \argmin_{0\leq \alpha\leq 1}\; \eta_2 \alpha^2
- 2\eta_1\alpha \label{eq:quad}.
\end{align}
where 
\begin{align*}
    \eta_2 &= c_1- 2c_3+ c_2 \\
    \eta_1 & = -c_3+c_2-c_4 \\
    c_1 &= \tr \widehat{Y}^\top {\Pi^*}^\top \widehat{L}_\mu \Pi^* \widehat{Y} \\
    c_2 &= \tr \widehat{Y}^\top D^\top \widehat{L}_\mu D \widehat{Y} \\
    c_3 &= \tr \widehat{Y}^\top D^\top \widehat{L}_\mu \Pi^* \widehat{Y} \\
    c_4 &= \tr \widetilde{Y}^\top\Pi_C^\top \widetilde{L} \Pi^* \widehat{Y}
\end{align*}
The following lemma, proof of which is quite straightforward, therefore omitted here, allows the efficient computation of Equation \eqref{eq:quad}:

\begin{lemma}
\label{lem:quad}
\begin{align*}
    \argmin_{0\leq \alpha\leq 1}\; \eta_2 \alpha^2
- 2\eta_1\alpha =\begin{cases}
\min\left(1, \frac{\eta_1}{\eta_2}\right) \quad &\eta_1\geq 0, \; \eta_2 >0, \\
1 \quad & \eta_1\geq 0, \; \eta_2 \leq 0, \\
0 \quad & \eta_1 < 0, \; \eta_2 \geq 0,\\
\ind\left\{\eta_1\leq \frac{1}{2}\eta_2 \right\} \quad & \eta_1 < 0, \; \eta_2 < 0. 
\end{cases}
\end{align*}
\end{lemma}
If $d_y<n$, which is usually the case in practice, we can greatly simplify the algorithm by changing the updated variable from $D$ to $D\widehat{Y}$. With this last change, we illustrate our GnCR algorithm for the shuffled regression problem in  Algorithm~\ref{GnCr General}.

\begin{algorithm2e}[ht]
\caption{Shuffled Linear Regression with GnCR}
   \label{GnCr General}
   \KwIn{Shuffled $Y = \Pi^{true}Y^{true} \in \mathbb{R}^{n\times d_y}$, $X \in \mathbb{R}^{n \times d_x}$, the seed permutation $\pi^{seeds}:C\to C^*$, continuation parameter $\gamma$, regularizer $\lambda$} 
\nl     $M = (X^\top X + \lambda \mathbf{I}_{d_x} )^{-1}$

\nl     $L =(X M X^\top - \mathbf{I})^2 + \lambda X M^\top M X^\top$ 

 \nl    $\widehat Y = Y[\overline C^*, :]\quad 
    \widetilde Y =  Y[C^*,:]$
    
\nl       $  \widehat L  = L[\overline C, \overline C] \quad  \widetilde L = L[C, \overline C] $

\nl     $\mu \leftarrow \mu_0$

\nl     $\widehat Y_{est} \leftarrow \frac{1}{n-|C|}\mathbf{1} \mathbf{1}^\top \widehat Y$

    \While{$\mu \leq \lambda_n(L)$}{
 \nl        $\widehat L_\mu \leftarrow \widehat L - \mu H$
 
     \nl $g_\mu(\widehat{Y}_{est}) \coloneqq 2\tr  \widetilde{Y}^\top \widetilde{L} \widehat{Y}_{est}  + \tr \widehat{Y}_{est}^\top  \widehat{L}_{\mu}  \widehat{Y}_{est}$
     
    \While{$\widehat Y_{est}$ has not converged}{
         \nl$ \Pi^* \leftarrow \argmin\limits_{\Pi} \tr  \left[\widehat{Y}( \widehat{Y}_{est}^\top  \widehat{L}_{\mu} + \widetilde{Y}^\top \widetilde{L} ) \Pi\right] $
         
         \nl$\widehat{Y}^* \leftarrow \Pi^* \widehat Y$
         
         \nl$\alpha \leftarrow \arg \min_{\alpha \in [0,1]} g_{\mu} (\alpha \widehat{Y}^* + (1-\alpha) \widehat Y_{est})$
         
         \nl Update: $\widehat Y_{est} \leftarrow \alpha \widehat{Y}^* + (1-\alpha) \widehat Y_{est}$
}
     \nl $\mu \leftarrow \gamma \mu$}
     \nl $Y_{est} = \text{collate}(\widehat {Y}_{est}, \widetilde{Y})$
     
     \nl $\beta_{est} = (X^\top X +\lambda\mathbf{I})^{-1} X^\top Y_{est}$
     
\KwOut{Estimates of $Y^{true}$ and the regression coefficient $\beta$}
\end{algorithm2e}

\subsection{Complexity bottleneck}\label{complexity}
Note, the complexity bottleneck in each iteration of the algorithm is the linear assignment problem in Step-9 of Algorithm~\ref{GnCr General}, which is exactly solvable by the Hungarian Algorithm in $O(|\overline C|^3)$ steps \cite{kuhn1955hungarian,munkres1957algorithms}. However, for the special case of $d_y=1$, it is more convenient to make use of the following identity,
\begin{align*}
    \argmin_{\Pi} \tr  \left[\widehat{Y}( \widehat{Y}_{est}^\top  \widehat{L}_{\mu} + \widetilde{Y}^\top \widetilde{L}_{\mu} ) \Pi\right]  = \argmin_{\Pi} \tr  \left[( \widehat{Y}_{est}^\top  \widehat{L}_{\mu} + \widetilde{Y}^\top \widetilde{L}_{\mu} ) \Pi \widehat{Y}\right] .
\end{align*}
Invoking the rearrangement inequality on the pre-sorted $\widehat{Y}$, as in the Step-2 of \refalg{Seriation_Algorithm}, the complexity of the assignment step can be further reduced to $O(|\overline C|\log |\overline C|)$, a dramatic cut in computational costs, which will let us apply our algorithm on large instances of real datasets without infeasible computational cost.       

\subsection{Comparison to other recent gradient descent methods}
Our algorithm, although similar to previous gradient descent algorithms for shuffled regression in \cite{xie2020hypergradient, zhang2021benefits}, differs from these works in important ways. In \cite{xie2020hypergradient}, authors also take the gradient based on the joint update of the estimate of planted permutation and the covariates ($\beta$). However the order of interplay between these two is reversed. That is, while we take the least square update on $\beta$ given the estimate of $\Pi^{true}$ as in \eqref{1241}, they consider the linear assignment update of $\Pi^{true}$ based on an estimate of $\beta$,  which leads to a more involved gradient computation compared to ours. \cite{zhang2021benefits}, on the other hand, uses \emph{projected} gradient descent, which involves a linear assignment problem of complexity $O(n^3)$, without the possibility of simplification we discussed in \refsec{complexity}. Our use of graduated non-convexity and integration of use of seeds is novel compared to both. 

\section{Numerical experiments}
\label{sec:experiments}
 In \refsec{sec:algorithms}, we describe the benchmark algorithms we consider in our numerical experiments, the choice of which was based on the code availability and ease of of implementation with the promise of competent performance. In \refsec{sec:metrics}, we define the metrics we use to evaluate the results. In \refsec{sec:synthetic}, we present results over synthetic datasets, and in \refsec{sec:real}, we compare the performances over real datasets, and elaborate on the role of seeds in the recovery. 
 
\subsection{Algorithms}
 \label{sec:algorithms}
In the next sections, we perform numerical experiments that compare our shuffled regression algorithm with the self-moment estimator from \cite{abid2017linear} and two other benchmark methods. 
\begin{itemize}
\item \textbf{Ordinary least squares regression}. We solve the least squares problem on the unshuffled data to provide a baseline of how good the linear fit is under the true permutation.
\item \textbf{Naive Alternative Optimization}.
    We implement an intuitive algorithm, proposed in \cite{haghighatshoar2017signal}, that iterates between solving ordinary linear regression (least squares), and finding the best permutation through solving linear assignment problem, until convergence, shown in \refalg{naive}. Note that the naive alternative optimization, although very efficient, is not expected to be as competent as other methods, since it does not take the optimization of $\beta$ and $\Pi$ together, and it is likely to get stuck in local optima. Our experimental results in the last section suggest that this is indeed the case. 
\end{itemize}

\begin{algorithm2e}[ht]
\caption{Naive Alternative Optimization for Shuffled Regression}
   \label{naive}
   \KwIn{Shuffled $Y = \Pi^{true}Y^{true} \in \mathbb{R}^{n\times d_y}$, $X \in \mathbb{R}^{n \times d_x}$, regularizer $\lambda$} 
   $\Pi_{est} = \Pi_0$; $\beta_{est} = \beta_0$ 
   
\While{$\beta_{est}$ has not converged}{
$\Pi_{est} = \argmin_{\Pi}\|\Pi Y - X \beta_{est}\|_F^2 $ \hspace{1em}\tcp{ \textcolor{magenta}{  via Hungarian Algorithm } }

$\beta_{est} =  \left( X^\top X + \lambda \mathbf{I}_{d_x} \right)^{-1}X^\top \Pi_{est} Y$ 
}
  
\KwOut{Estimates of $\Pi^{true}$ and the regression coefficients $\beta$ ($\Pi_{est}, \beta_{est}$)}
\end{algorithm2e}

\subsection{Metrics}
\label{sec:metrics}
     For each algorithm we report four different success measures. The first is the normalized overlap between the planted permutation, $\Pi^{true}$, and the estimate of the algorithm, that is $\langle \Pi_{est},\Pi^{true}\rangle/n$. This metric, obviously, measures the success in recovery of the planted permutation for algorithms. The second metric we use is the normalized correlation between the $\beta$ estimate of the algorithms, denoted $\beta_{est}$, and the \emph{true} $\beta$ of the generative model \eqref{gaus}, that is 
     \begin{align*}
       \frac{\langle \beta, \beta_{est}\rangle}{\|\beta\|\|\beta_{est}\|}.  
     \end{align*}
     In real datasets, due to lack of such natural definition of true coefficients, we take the output of ordinary linear regression problem \eqref{1241}, that is the case where $\Pi^{true}$ is known, as true $\beta$. The last two success metrics we report are the commonly used train and test errors between the true labels and the algorithm's estimate. Our train error calculation takes into account the algorithm's estimate of the planted shuffle and the scale of features, that is given by    
     \begin{align}\label{etrain}
        E_{\mathrm{train}} = \frac{\|\Pi_{est}Y-X\beta_{est}\|_F }{\|Y\|_F}.
    \end{align}
    Test error is calculated in the same manner, but for the test data we assume there is no shuffling. This reflects the nature of shuffled linear regression in common real scenarios where the publicized training data is obscured by shuffling but the modelling is conducted by another party, and therefore the goal is to ensure the \emph{learnability} without revealing the labels of individuals. 
    
\subsection{Experiments on Synthetic Datasets}
\label{sec:synthetic}
In this section, we compare and contrast the behavior of our algorithm with respect to the proposed baselines in a set of synthetic experiments. Across these experiments we generate data from a simple shuffled linear regression model \eqref{gaus} with $\Pi^{true}$ drawn uniformly at random from the set of  permutations of size $n$ and $\epsilon$ drawn i.i.d. from a Gaussian distribution with mean 0 and standard deviation $\sigma$.

\begin{figure}[ht]
    \includegraphics[width=1\linewidth]{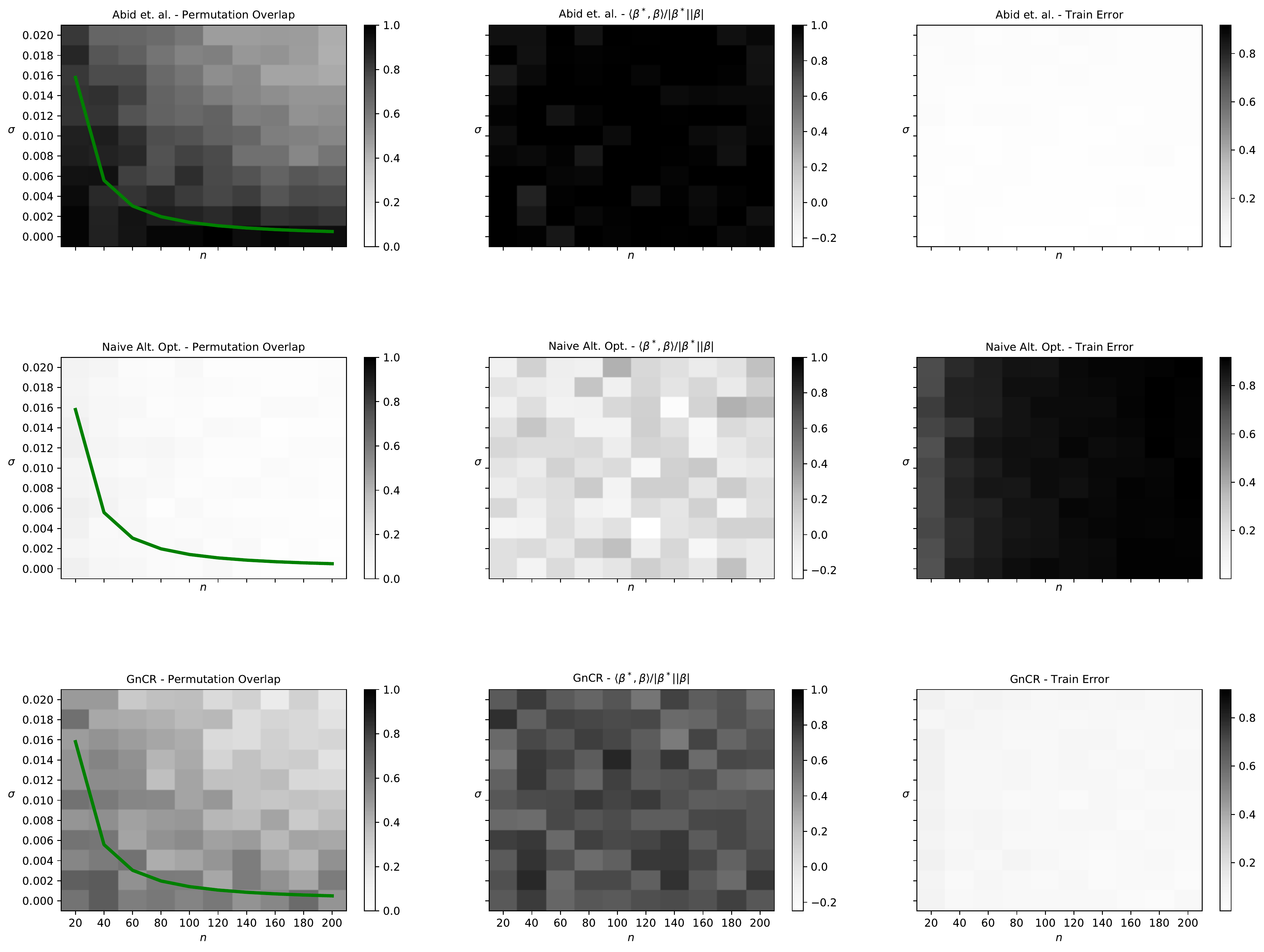}
    \centering
    \caption{Average performance of different algorithms for shuffled regression on synthetic data. For each choice of input size $n$ and noise level $\sigma$, we run 40 random experiments and plot the average performance in gray scale. Each experiment generates $n$ points at random in $\mathbb R^2$, a random regression coefficient vector in $\beta\in \mathbb R^{2\times 1}$ , and a random $n\times n$ permutation matrix $\Pi^{true}$. Each column reports the results for each of the algorithms described in Section \ref{sec:algorithms}. We show in gray scale (darker is higher) the average overlap with the planted permutation $\Pi^{true}$ (left column), the average correlation with the true regression coefficients (center column), and the train error defined in Section \ref{sec:metrics}.}
     \label{fig:n}
    \end{figure}

Throughout the numerical experiments, we fix the dimension parameters $d_x=2$ and $d_y=1$. For each $n\in\{20,40,\ldots,200\}$ and $\sigma\in\{0, 0.002,\ldots,0.02 \}$ we draw the entries of the matrix $X\in\bb R^{n\times d_x}$ i.i.d. with standard Gaussian distribution, i.e. $\mathcal{N}(0,1)$, and the entries of the noise vector $\epsilon$ i.i.d $\mathcal{N}(0,\sigma^2)$. After the observation matrix $Y$ is generated according to the model \eqref{gaus}, each algorithm is fed with the pair $X,Y$ and their output is recorded. This process has been independently repeated 40 times to get an average behavior, and each square in the grid plots of \reffig{fig:n} shows the average value of the metrics discussed previously across the repeated experiments. We do not report the test error in our synthetic dataset experiments as the normalized $\beta$ correlation naturally captures the extent to which every algorithm \emph{learns} the model already. In the \emph{permutation overlap} grid plots, information theoretic bound in \eqref{mltrue} for the perfect recovery (assuming $C_1=3$) is given by the green curve.     

    The numerical experiments we report in \reffig{fig:n} suggest some interesting observations. First, permutation recovery is extremely hard in this problem, as already implied by the unrealistically high SNR requirement given in \eqref{mltrue}. Even a very weak noise, compared to the input matrix $X$, is sufficient for the algorithms to fail in the recovery. The accuracy performance of our algorithm stays between the self-moment algorithm of \cite{abid2017linear} and the naive alternative optimization. Self-moment algorithm in this model seems to closely follow the information theoretic bound in permutation recovery, which would be an interesting theoretical question to explore in a future work.   

\subsection{Experiments on Real Datasets}
\label{sec:real}
We next conduct experiments on various real datasets from the UCI repository~\cite{Dua:2019}, and in \cite{data}. These datasets have a ground-truth correspondence between the data and labels. First, we randomly separate the dataset to training and test parts with the ratio of $4:1$. We shuffle the labels in the training set by applying a permutation drawn uniformly at random, and we run the shuffled regression algorithms on the same shuffled training data. Test error is calculated on the test data as in \eqref{etrain}, but without $\Pi_{est}$, as we assumed there is no shuffling in the test data.   

Since there is no guarantee that the linear model \eqref{gaus} holds in real datasets, we needed to pre-process the data before feeding it into our algorithm. First, some outliers are removed from the datasets to improve the linearity. Furthermore, note that different features in datasets might potentially have different mean value and/or range. The second pre-processing step accordingly was to ensure that features that included a negative value to be standardized to have 0-mean and unit standard deviation, and those with all non-negative values to be normalized to range between 0 and 1. The output of the algorithm (estimated $\beta$) is then re-scaled and offset to compensate for the pre-processing.    

We consider the same success measures referred to previously: permutation overlap, $\beta$ correlation, train, and test errors. We also report the computation time of each algorithm in seconds, as measured in our commercial laptop (Intel(R) Core(TM) i7-10510 CPU @ 1.8 GHz, 16 GB RAM), which are given for comparative purposes. Experiments on each dataset has been repeated ten times with random partition of the data to the train and test sets and the values in \reftab{real} report the average.

    
\begin{table}
\centering
{Ordinary Least Squares}
\begin{tabular}{lcc}
\hline
{\it dataset} & {\it Train Error} & {\it Test Error}\\
\hline
chloromethane ($n = 72$, $d_x=2$, $d_y=1$)             & 0.123       & 0.13          \\
NBA ($n = 404$, $d_x=3$, $d_y=1$)             & 0.068       & 0.065          \\
airfoil ($n = 1202$, $d_x=6$, $d_y=1$)             & 0.039       & 0.039          \\
rock strength ($n = 24$, $d_x=9$, $d_y=1$)             & 0.039       & 0.045          \\
concrete ($n = 824$, $d_x=9$, $d_y=1$)             & 0.262       & 0.262          \\
LPGA 2009 ($n = 116$, $d_x=13$, $d_y=1$)             & 0.034       & 0.041          \\
\hline
\end{tabular}
\\
\vspace{1.2em}

Self moments (\cite{abid2017linear})
\begin{tabular}{lccccc}
\hline
{\it dataset} & {\it Perm. Ov.} & {$\beta$ \it Corr.} & {\it Train E.} & {\it Test E.} & {\it Time}\\
\hline
chloromethane & 0.122 &1.0 &0.111 &0.13 &0.01 \\
NBA &0.06 &0.855 &0.007 &0.097 &0.309 \\ 
airfoil &0.001 &-0.043 &0.001 &0.083 &13.017 \\
rock strength &0.025 &-0.051 &0.005 &0.142 &10.755 \\
concrete &0.001 &-0.076 &0.011 &0.5 &32.392 \\
LPGA 2009 &0.01 &0.248 &0.004 &0.167 &25.449 \\
\hline
\end{tabular}
\\
\vspace{1.2em}

Naive Alternative Optimization (\refalg{naive})
\begin{tabular}{lccccc}
\hline
{\it dataset} & {\it Perm. Ov.} & {$\beta$ \it Corr.} & {\it Train E.} & {\it Test E.} & {\it Time}\\
\hline
chloromethane &0.06  &0.0 &0.67 &0.702 &0.001 \\
NBA &0.006 &0.592 &0.111 &0.116 &0.001 \\ 
airfoil &0.001 &-0.49 &0.052 &0.056 &0.001 \\
rock strength &0.062 &0.73 &0.058 &0.097 &0.001 \\
concrete &0.002 &0.303 &0.382 &0.417 &0.001 \\
LPGA 2009 &0.005 &0.147 &0.085 &0.119 &0.003 \\
\hline
\end{tabular}
\\
\vspace{1.2em}

GnCR (\refalg{GnCr General})
\begin{tabular}{lccccc}
\hline
{\it dataset} & {\it Perm. Ov.} & {$\beta$ \it Corr.} & {\it Train E.} & {\it Test E.} & {\it Time}\\
\hline
chloromethane &0.122 &1.0 &0.122 &0.144 &0.001 \\
NBA &0.008 &0.807 &0.019 &0.096 &2.4 \\ 
airfoil &0.0 &0.011 &0.004 &0.098 &12.885 \\
rock strength &0.221 &0.432 &0.01 &0.054 &0.078 \\
concrete &0.001 &0.207 &0.031 &0.607 &20.523 \\
LPGA 2009 &0.032 &0.893 &0.008 &0.057 &0.293 \\
\hline
\end{tabular}

\caption{
    We evaluate the performance of the shuffled regression algorithms described in Section \ref{sec:algorithms} on different real datasets from \cite{Dua:2019,data}, using the metrics described in Section \ref{sec:metrics}. The datasets belong to, in the order of appearance: the chloromethane peak ratio and concentration in water \cite{lavagnini2007statistical}; height, weight and age of NBA players in the 2013-2014 season; aerodynamic and acoustic tests of airfoil blade sections by NASA; compressive strength of rock specimens with 8 predictors \cite{ali2014empirical}; compressive strength of concrete sample \cite{yeh1998modeling}; and lastly the performance statistics of LPGA golfers in 2009. The first table corresponds to ordinary linear regression on the unshuffled data, and it provides a useful baseline for comparison, with the dimensions of each dataset shown in parantheses. 
    }\label{real}
\end{table}
    
    From \reftab{real}, we can make some observations as follows. Our GnCR algorithm, in contrast to the case of Gaussian generative model discussed previously, seems to be the most successful algorithm of all considered. It also is observed to be faster than the self-moment algorithm for most datasets. Nevertheless, permutation recovery again stands out to be a much harder task than finding a correlated coefficient vector ($\beta$), except in very small datasets. This, however, as we note is not a shortcoming of the algorithms but a manifestation of the information theoretic hardness of the problem as the train error values suggest. Relative success of GnCR algorithm in coefficient recovery compared to the permutation recovery highlights the promising suitability of GnCR algorithm in realistic anonymous learning applications.       
    
\begin{figure}[ht]
    \includegraphics[width=1.0\linewidth]{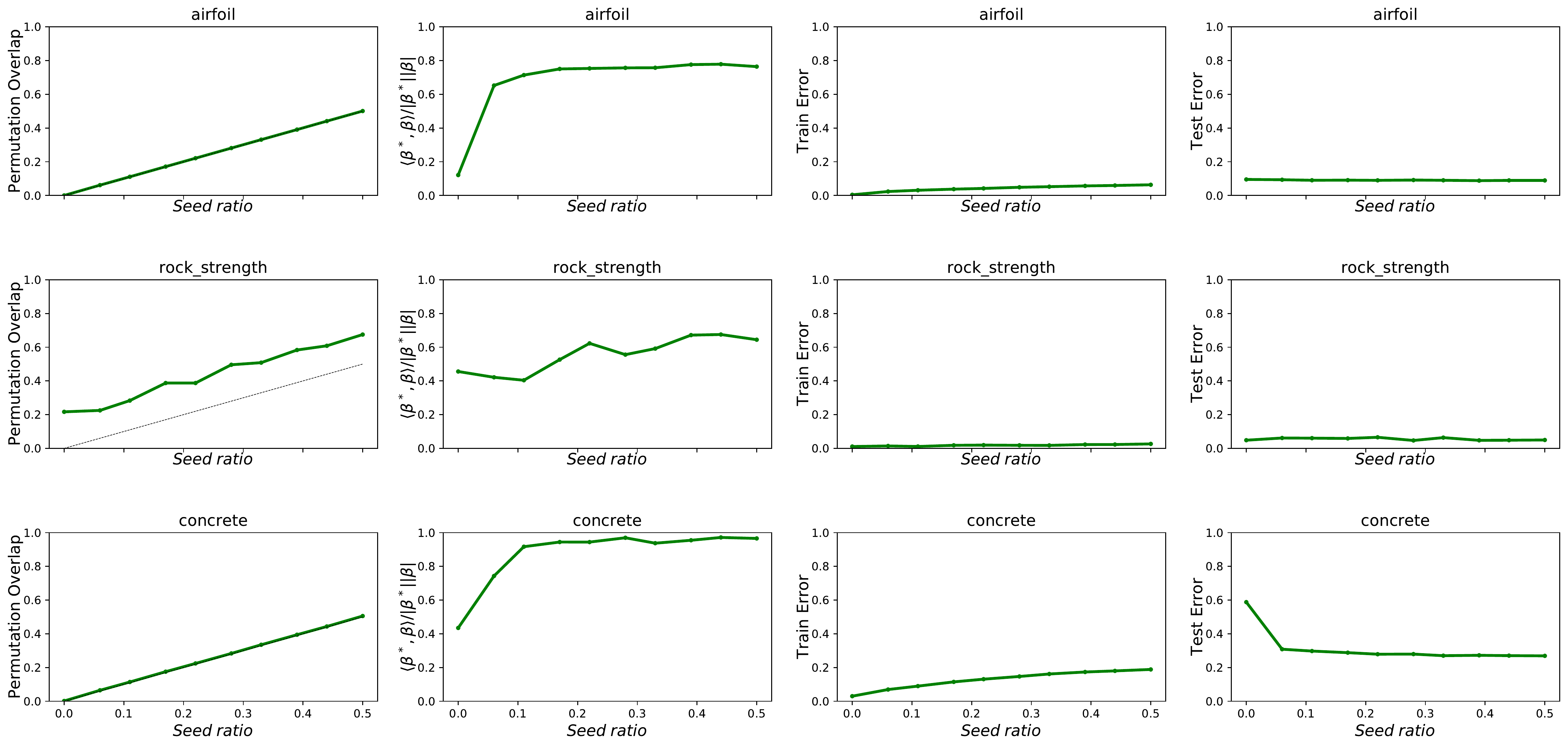}
    \centering
    \caption{Results on real datasets with provided seeds.
    We evaluate the performance of our GnCR algorithm with varying seed ratios on the real datasets for which the GnCR showed the worst performance in the unseeded case. Note that the permutation overlap includes the given seeds, accordingly the diagonal ($x=y$) line in the permutation recovery plots represent the baseline. 
    }
    \label{seeds}
    \end{figure}

    \emph{Effect of partial information.}
    As discussed before, our algorithm is designed to leverage partial information (seeds), if available, for the recovery of $\Pi^{true}$ and the regression coefficients. We define the \emph{seed ratio} to be the ratio of the number of randomly chosen seeds to total number of samples in the training set.
    We chose the three real datasets that the GnCR algorithm showed poorer performance previously, in terms of coefficient recovery, to show how side information helps our algorithm. On these three datasets, we report the plots of our success metrics for differing seed ratio in \reffig{seeds}. Each setting for each parameter is repeated ten times with randomly selected seeds.
    
    The figure shows our algorithm to improve in recovering the regression coefficients as a result of incorporating side information, although surprisingly the permutation recovery is not noticeably affected. Train error increases in some cases with the seeds which suggest their role in ameliorating the \emph{overfitting} observed previously in experiments on real datasets without the seed information. These observations suggest a few seeds are very useful in rendering the data \emph{learnable} without sacrificing the anonimity of the rest, and therefore further study of algorithms that exploit the seed information could be valuable in the practical applications of the shuffled regression problem. 
    
\section{Discussion}
\label{sec:discussion}

Motivated by the connection between shuffled linear regression and seriation, in this paper we propose a new algorithm for the former through a Bayesian formulation of the problem. 
 We perform extensive numerical experiments to test our algorithm and compare it to the selected benchmark methods.  

Our experiments suggest that our algorithm has comparable performance with state-of-the-art empirical methods. One advantage that our algorithm presents over existing approaches is that it naturally generalizes to the multi-output setting, while also being faster. According to the experiments in real datasets, it seems to stand on the \emph{sweet spot} between efficiency and accuracy. Furthermore, we devised our algorithm so that it incorporates the side information in the form of seeds, if available. We showed that this feature can be a significant advantage compared to other algorithms, given that the permutation recovery is information theoretically impossible in several real scenarios without the seeds. 



\section*{Acknowledgments}
The authors would like to thank Bianca Dumitrascu for valuable discussions, particularly regarding the initial conceptualization of the problem.

\bibliographystyle{ba}
\bibliography{refs}

\begin{thebibliography}{57}
\newcommand{\enquote}[1]{``#1''}
\expandafter\ifx\csname natexlab\endcsname\relax\def\natexlab#1{#1}\fi
\expandafter\ifx\csname url\endcsname\relax
  \def\url#1{{\tt #1}}\fi
\expandafter\ifx\csname urlprefix\endcsname\relax\def\urlprefix{URL }\fi
\ifx\endbibitem\undefined \let\endbibitem\relax\fi

\bibitem[{Abid et~al.(2017)Abid, Poon, and Zou}]{abid2017linear}
Abid, A., Poon, A., and Zou, J. (2017).
\newblock \enquote{Linear regression with shuffled labels.}
\newblock {\em arXiv preprint arXiv:1705.01342\/}.
\endbibitem

\bibitem[{Ali et~al.(2014)Ali, Guang, and Ibrahim}]{ali2014empirical}
Ali, E., Guang, W., and Ibrahim, A. (2014).
\newblock \enquote{Empirical relations between compressive strength and
  microfabric properties of amphibolites using multivariate regression, fuzzy
  inference and neural networks: A comparative study.}
\newblock {\em Engineering geology\/}, 183: 230--240.
\endbibitem

\bibitem[{Atkins et~al.(1998)Atkins, Boman, and
  Hendrickson}]{atkins1998spectral}
Atkins, J.~E., Boman, E.~G., and Hendrickson, B. (1998).
\newblock \enquote{A spectral algorithm for seriation and the consecutive ones
  problem.}
\newblock {\em SIAM Journal on Computing\/}, 28(1): 297--310.
\endbibitem

\bibitem[{Baron et~al.(2019)Baron, Barve, Muraro, van~der Linden,
  Dharmadhikari, Lyubimova, de~Koning, and van Oudenaarden}]{baron2019cell}
Baron, C.~S., Barve, A., Muraro, M.~J., van~der Linden, R., Dharmadhikari, G.,
  Lyubimova, A., de~Koning, E.~J., and van Oudenaarden, A. (2019).
\newblock \enquote{Cell type purification by single-cell transcriptome-trained
  sorting.}
\newblock {\em Cell\/}, 179(2): 527--542.
\endbibitem

\bibitem[{Bonawitz et~al.(2017)Bonawitz, Ivanov, Kreuter, Marcedone, McMahan,
  Patel, Ramage, Segal, and Seth}]{bonawitz2017practical}
Bonawitz, K., Ivanov, V., Kreuter, B., Marcedone, A., McMahan, H.~B., Patel,
  S., Ramage, D., Segal, A., and Seth, K. (2017).
\newblock \enquote{Practical secure aggregation for privacy-preserving machine
  learning.}
\newblock In {\em proceedings of the 2017 ACM SIGSAC Conference on Computer and
  Communications Security\/}, 1175--1191.
\endbibitem

\bibitem[{Cai et~al.(2021)Cai, Wang, and Zhang}]{cai2021cost}
Cai, T.~T., Wang, Y., and Zhang, L. (2021).
\newblock \enquote{The cost of privacy: Optimal rates of convergence for
  parameter estimation with differential privacy.}
\newblock {\em The Annals of Statistics\/}, 49(5): 2825--2850.
\endbibitem

\bibitem[{Carpentier and Schl{\"u}ter(2016)}]{carpentier2016learning}
Carpentier, A. and Schl{\"u}ter, T. (2016).
\newblock \enquote{Learning relationships between data obtained independently.}
\newblock In {\em Artificial Intelligence and Statistics\/}, 658--666.
\endbibitem

\bibitem[{Collier and Dalalyan(2016)}]{collier2016minimax}
Collier, O. and Dalalyan, A.~S. (2016).
\newblock \enquote{Minimax rates in permutation estimation for feature
  matching.}
\newblock {\em The Journal of Machine Learning Research\/}, 17(1): 162--192.
\endbibitem

\bibitem[{Crandall and Huttenlocher(2006)}]{crandall2006weakly}
Crandall, D.~J. and Huttenlocher, D.~P. (2006).
\newblock \enquote{Weakly supervised learning of part-based spatial models for
  visual object recognition.}
\newblock In {\em European conference on computer vision\/}, 16--29. Springer.
\endbibitem

\bibitem[{Cullina and Kiyavash(2016)}]{cullina2016improved}
Cullina, D. and Kiyavash, N. (2016).
\newblock \enquote{Improved achievability and converse bounds for
  Erdos-R{\'e}nyi graph matching.}
\newblock In {\em Proceedings of the 2016 ACM SIGMETRICS International
  Conference on Measurement and Modeling of Computer Science\/}, 63--72. ACM.
\endbibitem

\bibitem[{Dua and Graff(2017)}]{Dua:2019}
Dua, D. and Graff, C. (2017).
\newblock \enquote{{UCI} Machine Learning Repository.}
\newline\urlprefix\url{http://archive.ics.uci.edu/ml}
\endbibitem

\bibitem[{Evangelopoulos et~al.(2018)Evangelopoulos, Brockmeier, Mu, and
  Goulermas}]{evangelopoulos2018continuation}
Evangelopoulos, X., Brockmeier, A.~J., Mu, T., and Goulermas, J.~Y. (2018).
\newblock \enquote{Continuation methods for approximate large scale object
  sequencing.}
\newblock {\em Machine Learning\/}, 1--32.
\endbibitem

\bibitem[{Fogel et~al.(2013)Fogel, Jenatton, Bach, and
  d'Aspremont}]{fogel2013convex}
Fogel, F., Jenatton, R., Bach, F., and d'Aspremont, A. (2013).
\newblock \enquote{Convex relaxations for permutation problems.}
\newblock In {\em Advances in Neural Information Processing Systems\/},
  1016--1024.
\endbibitem

\bibitem[{George and Pothen(1997)}]{george1997analysis}
George, A. and Pothen, A. (1997).
\newblock \enquote{An analysis of spectral envelope reduction via quadratic
  assignment problems.}
\newblock {\em SIAM Journal on Matrix Analysis and Applications\/}, 18(3):
  706--732.
\endbibitem

\bibitem[{Goel(1975)}]{goel1975re}
Goel, P.~K. (1975).
\newblock \enquote{On re-pairing observations in a broken random sample.}
\newblock {\em The Annals of Statistics\/}, 1364--1369.
\endbibitem

\bibitem[{Goemans(2015)}]{goemans2015smallest}
Goemans, M.~X. (2015).
\newblock \enquote{Smallest compact formulation for the permutahedron.}
\newblock {\em Mathematical Programming\/}, 153(1): 5--11.
\endbibitem

\bibitem[{Haghighatshoar and Caire(2017)}]{haghighatshoar2017signal}
Haghighatshoar, S. and Caire, G. (2017).
\newblock \enquote{Signal recovery from unlabeled samples.}
\newblock {\em IEEE Transactions on Signal Processing\/}, 66(5): 1242--1257.
\endbibitem

\bibitem[{Hardy et~al.(1952)Hardy, Littlewood, and
  Polya}]{hardy1952inequalities}
Hardy, G., Littlewood, J., and Polya, G. (1952).
\newblock \enquote{Inequalities cambridge univ.}
\newblock {\em Press, Cambridge\/}.
\endbibitem

\bibitem[{Hejblum et~al.(2019)Hejblum, Weber, Liao, Palmer, Churchill, Shadick,
  Szolovits, Murphy, Kohane, and Cai}]{hejblum2019probabilistic}
Hejblum, B.~P., Weber, G.~M., Liao, K.~P., Palmer, N.~P., Churchill, S.,
  Shadick, N.~A., Szolovits, P., Murphy, S.~N., Kohane, I.~S., and Cai, T.
  (2019).
\newblock \enquote{Probabilistic record linkage of de-identified research
  datasets with discrepancies using diagnosis codes.}
\newblock {\em Scientific data\/}, 6: 180298.
\endbibitem

\bibitem[{Herzenberg et~al.(2006)Herzenberg, Tung, Moore, Herzenberg, and
  Parks}]{herzenberg2006interpreting}
Herzenberg, L.~A., Tung, J., Moore, W.~A., Herzenberg, L.~A., and Parks, D.~R.
  (2006).
\newblock \enquote{Interpreting flow cytometry data: a guide for the
  perplexed.}
\newblock {\em Nature immunology\/}, 7(7): 681--685.
\endbibitem

\bibitem[{Hsu et~al.(2017)Hsu, Shi, and Sun}]{hsu2017linear}
Hsu, D.~J., Shi, K., and Sun, X. (2017).
\newblock \enquote{Linear regression without correspondence.}
\newblock In {\em Advances in Neural Information Processing Systems\/},
  1531--1540.
\endbibitem

\bibitem[{Kazemi et~al.(2015)Kazemi, Hassani, and
  Grossglauser}]{kazemi2015growing}
Kazemi, E., Hassani, S.~H., and Grossglauser, M. (2015).
\newblock \enquote{Growing a graph matching from a handful of seeds.}
\newblock {\em Proceedings of the VLDB Endowment\/}, 8(10): 1010--1021.
\endbibitem

\bibitem[{Kuhn(1955)}]{kuhn1955hungarian}
Kuhn, H.~W. (1955).
\newblock \enquote{The Hungarian method for the assignment problem.}
\newblock {\em Naval research logistics quarterly\/}, 2(1-2): 83--97.
\endbibitem

\bibitem[{Lavagnini and Magno(2007)}]{lavagnini2007statistical}
Lavagnini, I. and Magno, F. (2007).
\newblock \enquote{A statistical overview on univariate calibration, inverse
  regression, and detection limits: Application to gas chromatography/mass
  spectrometry technique.}
\newblock {\em Mass spectrometry reviews\/}, 26(1): 1--18.
\endbibitem

\bibitem[{Li(2015)}]{li2015ensuring}
Li, J. (2015).
\newblock \enquote{Ensuring privacy in a personal health record system.}
\newblock {\em Computer\/}, 48(2): 24--31.
\endbibitem

\bibitem[{Li and Shaw(2004)}]{li2004protection}
Li, J. and Shaw, M.~J. (2004).
\newblock \enquote{Protection of health information in data mining.}
\newblock {\em International Journal of Healthcare Technology and
  Management\/}, 6(2): 210--222.
\endbibitem

\bibitem[{Lim and Wright(2014)}]{lim2014beyond}
Lim, C.~H. and Wright, S. (2014).
\newblock \enquote{Beyond the birkhoff polytope: Convex relaxations for vector
  permutation problems.}
\newblock In {\em Advances in Neural Information Processing Systems\/},
  2168--2176.
\endbibitem

\bibitem[{Lowe(2004)}]{lowe2004distinctive}
Lowe, D.~G. (2004).
\newblock \enquote{Distinctive image features from scale-invariant keypoints.}
\newblock {\em International journal of computer vision\/}, 60(2): 91--110.
\endbibitem

\bibitem[{Lyzinski et~al.(2016)Lyzinski, Fishkind, Fiori, Vogelstein, Priebe,
  and Sapiro}]{lyzinski2016graph}
Lyzinski, V., Fishkind, D.~E., Fiori, M., Vogelstein, J.~T., Priebe, C.~E., and
  Sapiro, G. (2016).
\newblock \enquote{Graph matching: Relax at your own risk.}
\newblock {\em {IEEE} Trans. Pattern Anal. Mach. Intell.\/}, 38(1): 60--73.
\endbibitem

\bibitem[{Lyzinski et~al.(2014)Lyzinski, Fishkind, and
  Priebe}]{lyzinski2014seeded}
Lyzinski, V., Fishkind, D.~E., and Priebe, C.~E. (2014).
\newblock \enquote{Seeded graph matching for correlated Erd{\"o}s-R{\'e}nyi
  graphs.}
\newblock {\em The Journal of Machine Learning Research\/}, 15(1): 3513--3540.
\endbibitem

\bibitem[{Munkres(1957)}]{munkres1957algorithms}
Munkres, J. (1957).
\newblock \enquote{Algorithms for the assignment and transportation problems.}
\newblock {\em Journal of the society for industrial and applied
  mathematics\/}, 5(1): 32--38.
\endbibitem

\bibitem[{Narayanan and Shmatikov(2008)}]{narayanan2008robust}
Narayanan, A. and Shmatikov, V. (2008).
\newblock \enquote{Robust de-anonymization of large sparse datasets.}
\newblock In {\em IEEE Symposium on Security and Privacy\/}, 111--125.
\endbibitem

\bibitem[{Onaran et~al.(2016)Onaran, Garg, and Erkip}]{onaran2016optimal}
Onaran, E., Garg, S., and Erkip, E. (2016).
\newblock \enquote{Optimal de-anonymization in random graphs with community
  structure.}
\newblock In {\em Signals, Systems and Computers, 2016 50th Asilomar Conference
  on\/}, 709--713. IEEE.
\endbibitem

\bibitem[{Onaran and Villar(2017)}]{onaran2017projected}
Onaran, E. and Villar, S. (2017).
\newblock \enquote{Projected power iteration for network alignment.}
\newblock In {\em Wavelets and Sparsity XVII\/}, volume 10394, 103941C.
  International Society for Optics and Photonics.
\endbibitem

\bibitem[{Pananjady et~al.(2017{\natexlab{a}})Pananjady, Wainwright, and
  Courtade}]{pananjady2017denoising}
Pananjady, A., Wainwright, M.~J., and Courtade, T.~A. (2017{\natexlab{a}}).
\newblock \enquote{Denoising linear models with permuted data.}
\newblock In {\em Information Theory (ISIT), 2017 IEEE International Symposium
  on\/}, 446--450. IEEE.
\endbibitem

\bibitem[{Pananjady et~al.(2017{\natexlab{b}})Pananjady, Wainwright, and
  Courtade}]{pananjady2017linear}
--- (2017{\natexlab{b}}).
\newblock \enquote{Linear regression with shuffled data: Statistical and
  computational limits of permutation recovery.}
\newblock {\em IEEE Transactions on Information Theory\/}, 64(5): 3286--3300.
\endbibitem

\bibitem[{Peng and Tsakiris(2020)}]{peng2020linear}
Peng, L. and Tsakiris, M.~C. (2020).
\newblock \enquote{Linear regression without correspondences via concave
  minimization.}
\newblock {\em IEEE Signal Processing Letters\/}, 27: 1580--1584.
\endbibitem

\bibitem[{Recanati et~al.(2018)Recanati, Servant, Vert, and
  d'Aspremont}]{recanati2018robust}
Recanati, A., Servant, N., Vert, J.-P., and d'Aspremont, A. (2018).
\newblock \enquote{Robust Seriation and Applications to Cancer Genomics.}
\newblock {\em arXiv preprint arXiv:1806.00664\/}.
\endbibitem

\bibitem[{Rigollet and Weed(2019)}]{rigollet2019uncoupled}
Rigollet, P. and Weed, J. (2019).
\newblock \enquote{Uncoupled isotonic regression via minimum Wasserstein
  deconvolution.}
\newblock {\em Information and Inference: A Journal of the IMA\/}, 8(4):
  691--717.
\endbibitem

\bibitem[{Robinson(1951)}]{robinson1951method}
Robinson, W.~S. (1951).
\newblock \enquote{A method for chronologically ordering archaeological
  deposits.}
\newblock {\em American antiquity\/}, 16(4): 293--301.
\endbibitem

\bibitem[{Rose and Mian(2014)}]{rose2014signaling}
Rose, C. and Mian, I.~S. (2014).
\newblock \enquote{Signaling with identical tokens: Upper bounds with energy
  constraints.}
\newblock In {\em 2014 IEEE International Symposium on Information Theory\/},
  1817--1821. IEEE.
\endbibitem

\bibitem[{Sahni and Gonzalez(1976)}]{sahni1976p}
Sahni, S. and Gonzalez, T. (1976).
\newblock \enquote{P-complete approximation problems.}
\newblock {\em Journal of the ACM (JACM)\/}, 23(3): 555--565.
\endbibitem

\bibitem[{Shi et~al.(2021)Shi, Li, and Cai}]{shi2021spherical}
Shi, X., Li, X., and Cai, T. (2021).
\newblock \enquote{Spherical regression under mismatch corruption with
  application to automated knowledge translation.}
\newblock {\em Journal of the American Statistical Association\/}, 116(536):
  1953--1964.
\endbibitem

\bibitem[{Shirani et~al.(2017)Shirani, Garg, and Erkip}]{shirani2017seeded}
Shirani, F., Garg, S., and Erkip, E. (2017).
\newblock \enquote{Seeded graph matching: Efficient algorithms and theoretical
  guarantees.}
\newblock In {\em 2017 51st Asilomar Conference on Signals, Systems, and
  Computers\/}, 253--257. IEEE.
\endbibitem

\bibitem[{Slawski et~al.(2020{\natexlab{a}})Slawski, Ben-David, and
  Li}]{slawski2020two}
Slawski, M., Ben-David, E., and Li, P. (2020{\natexlab{a}}).
\newblock \enquote{Two-Stage Approach to Multivariate Linear Regression with
  Sparsely Mismatched Data.}
\newblock {\em J. Mach. Learn. Res.\/}, 21(204): 1--42.
\endbibitem

\bibitem[{Slawski et~al.(2019)Slawski, Ben-David et~al.}]{slawski2019linear}
Slawski, M., Ben-David, E., et~al. (2019).
\newblock \enquote{Linear regression with sparsely permuted data.}
\newblock {\em Electronic Journal of Statistics\/}, 13(1): 1--36.
\endbibitem

\bibitem[{Slawski et~al.(2020{\natexlab{b}})Slawski, Rahmani, and
  Li}]{slawski2020sparse}
Slawski, M., Rahmani, M., and Li, P. (2020{\natexlab{b}}).
\newblock \enquote{A sparse representation-based approach to linear regression
  with partially shuffled labels.}
\newblock In {\em Uncertainty in Artificial Intelligence\/}, 38--48. PMLR.
\endbibitem

\bibitem[{Stachniss et~al.(2016)Stachniss, Leonard, and
  Thrun}]{stachniss2016simultaneous}
Stachniss, C., Leonard, J.~J., and Thrun, S. (2016).
\newblock \enquote{Simultaneous localization and mapping.}
\newblock In {\em Springer Handbook of Robotics\/}, 1153--1176. Springer.
\endbibitem

\bibitem[{Steorts et~al.(2016)Steorts, Hall, and
  Fienberg}]{steorts2016bayesian}
Steorts, R.~C., Hall, R., and Fienberg, S.~E. (2016).
\newblock \enquote{A Bayesian approach to graphical record linkage and
  deduplication.}
\newblock {\em Journal of the American Statistical Association\/}, 111(516):
  1660--1672.
\endbibitem

\bibitem[{Tsakiris et~al.(2020)Tsakiris, Peng, Conca, Kneip, Shi, and
  Choi}]{tsakiris2020algebraic}
Tsakiris, M.~C., Peng, L., Conca, A., Kneip, L., Shi, Y., and Choi, H. (2020).
\newblock \enquote{An algebraic-geometric approach for linear regression
  without correspondences.}
\newblock {\em IEEE Transactions on Information Theory\/}.
\endbibitem

\bibitem[{Unnikrishnan et~al.(2018)Unnikrishnan, Haghighatshoar, and
  Vetterli}]{unnikrishnan2018unlabeled}
Unnikrishnan, J., Haghighatshoar, S., and Vetterli, M. (2018).
\newblock \enquote{Unlabeled sensing with random linear measurements.}
\newblock {\em IEEE Transactions on Information Theory\/}, 64(5): 3237--3253.
\endbibitem

\bibitem[{Winner(2009)}]{data}
Winner, L. (2009).
\newblock \enquote{Miscellaneous Datasets.}
\newblock \url{http://users.stat.ufl.edu/~winner/datasets.html}.
\newblock Accessed: 2020-09-05.
\endbibitem

\bibitem[{Xie et~al.(2020)Xie, Mao, Zuo, Xu, Ye, Zhao, and
  Zha}]{xie2020hypergradient}
Xie, Y., Mao, Y., Zuo, S., Xu, H., Ye, X., Zhao, T., and Zha, H. (2020).
\newblock \enquote{A hypergradient approach to robust regression without
  correspondence.}
\newblock {\em arXiv preprint arXiv:2012.00123\/}.
\endbibitem

\bibitem[{Yartseva and Grossglauser(2013)}]{yartseva2013performance}
Yartseva, L. and Grossglauser, M. (2013).
\newblock \enquote{On the performance of percolation graph matching.}
\newblock In {\em Proceedings of the first ACM conference on Online social
  networks\/}, 119--130.
\endbibitem

\bibitem[{Yeh(1998)}]{yeh1998modeling}
Yeh, I.-C. (1998).
\newblock \enquote{Modeling of strength of high-performance concrete using
  artificial neural networks.}
\newblock {\em Cement and Concrete research\/}, 28(12): 1797--1808.
\endbibitem

\bibitem[{Zhang et~al.(2019)Zhang, Slawski, and Li}]{zhang2019permutation}
Zhang, H., Slawski, M., and Li, P. (2019).
\newblock \enquote{Permutation recovery from multiple measurement vectors in
  unlabeled sensing.}
\newblock In {\em 2019 IEEE International Symposium on Information Theory
  (ISIT)\/}, 1857--1861. IEEE.
\endbibitem

\bibitem[{Zhang et~al.(2021)Zhang, Slawski, and Li}]{zhang2021benefits}
--- (2021).
\newblock \enquote{The benefits of diversity: Permutation recovery in unlabeled
  sensing from multiple measurement vectors.}
\newblock {\em IEEE Transactions on Information Theory\/}.
\endbibitem

\end{thebibliography}


\end{document}